\documentclass[12pt,draft]{article} 
\usepackage{amsmath}
\usepackage{amssymb}
\usepackage{dsfont}
\usepackage{amsthm}
   \font\twelvebm                       = cmmib10 at 12truept
   \font\tenbm                          = cmmib10 at 10truept
   \font\sevenbm                        = cmmib10 at 7truept
  at 10truept  at 10truept  at
10truept  at 10truept
 at 10truept

\mathchardef \BGamma            = "0900 \mathchardef \BDelta =
"0901 \mathchardef \BTheta            = "0902 \mathchardef
\BLambda           = "0903 \mathchardef \BXi               = "0904
\mathchardef \BPi               = "0905 \mathchardef \BSigma =
"0906 \mathchardef \BUpsilon          = "0907 \mathchardef \BPhi =
"0908 \mathchardef \BPsi              = "0909 \mathchardef \BOmega
= "090A \mathchardef \Balpha            = "090B \mathchardef
\Bbeta             = "090C \mathchardef \Bgamma = "090D
\mathchardef \Bdelta            = "090E \mathchardef \Bepsilon =
"090F \mathchardef \Bzeta             = "0910 \mathchardef \Beta =
"0911 \mathchardef \Btheta = "0912 \mathchardef \Biota = "0913
\mathchardef \Bkappa            = "0914 \mathchardef \Blambda
= "0915 \mathchardef \Bmu               = "0916 \mathchardef \Bnu
= "0917 \mathchardef \Bxi               = "0918 \mathchardef \Bpi
= "0919 \mathchardef \Brho              = "091A \mathchardef
\Bsigma            = "091B \mathchardef \Btau = "091C \mathchardef
\Bupsilon          = "091D \mathchardef \Bphi = "091E \mathchardef
\Bchi              = "091F \mathchardef \Bpsi = "0920 \mathchardef
\Bomega            = "0921 \mathchardef \Bvarepsilon       = "0922
\mathchardef \Bvartheta         = "0923 \mathchardef \Bvarpi
= "0924 \mathchardef \Bvarrho = "0925 \mathchardef \Bvarsigma
= "0926 \mathchardef \Bvarphi           = "0927
\mathchardef \bA        = "0941 \mathchardef \bB        = "0942
\mathchardef \bC        = "0943 \mathchardef \bD        = "0944
\mathchardef \bE        = "0945 \mathchardef \bF        = "0946
\mathchardef \bG        = "0947 \mathchardef \bH        = "0948
\mathchardef \bI        = "0949 \mathchardef \bJ        = "094A
\mathchardef \bK        = "094B \mathchardef \bL        = "094C
\mathchardef \bM        = "094D \mathchardef \bN        = "094E
\mathchardef \bO        = "094F \mathchardef \bP        = "0950
\mathchardef \bQ        = "0951 \mathchardef \bR        = "0952
\mathchardef \bS        = "0953 \mathchardef \bT        = "0954
\mathchardef \bU        = "0955 \mathchardef \bV        = "0956
\mathchardef \bW        = "0957 \mathchardef \bX        = "0958
\mathchardef \bY        = "0959 \mathchardef \bZ        = "095A
\mathchardef \ba        = "0961 \mathchardef \bb        = "0962
\mathchardef \bc        = "0963 \mathchardef \bd        = "0964
\mathchardef \bee       = "0965 
\mathchardef \bff       = "0966 \mathchardef \bg        = "0967
\mathchardef \bh        = "0968
\mathchardef \bj        = "096A \mathchardef \bk        = "096B
\mathchardef \bl        = "096C \mathchardef \bm        = "096D
\mathchardef \bn        = "096E \mathchardef \bo        = "096F
\mathchardef \bp        = "0970 \mathchardef \bq        = "0971
\mathchardef \br        = "0972 \mathchardef \bs        = "0973
\mathchardef \bt        = "0974 \mathchardef \bu        = "0975
\mathchardef \bv        = "0976 \mathchardef \bw        = "0977
\mathchardef \bx        = "0978 \mathchardef \by        = "0979
\mathchardef \bz        = "097A

\font\tencb            = cmssbx10 scaled \magstep4 \font\eigcb =
cmssbx10 scaled \magstep2 \textfont8             = \tencb
\mathchardef\bAs       = "1841
\def\Asem#1#2{\mathop{\vrule height10.5pt depth5.5pt width0pt\bAs}_{#1}^{#2}}
\def\asem#1#2{
          \ifmmode
         \ifinner
            \raise0.9pt\hbox{$\scriptstyle\bAs$}_{#1}^{#2}
         \else
            \Asem{#1}{#2}
         \fi
          \fi
          }

\newtheorem{theo}{\small\bf Theorem}[section]

\newtheorem{lem}{\small\bf Lemma}[section]

\newtheorem{prop}{\small\bf Proposition} 
\newenvironment{PROP}{\begin{prop} \rm}{\end{prop}}
\newtheorem{rem}{\small\bf Remark}[section]

\newtheorem{defi}{\small\bf Definition}[section]

\newtheorem{cor}{\small\bf Corollary}[section]

\newtheorem{example}{\small\bf Example}[section]

\newenvironment{Proof}{{\small\bf Proof:}}{}
\renewcommand{\Pr}{\mathds{P}}

\newcommand{\be}{\begin{equation}}
\newcommand{\ee}{\end{equation}}

\newcommand{\R}{\mathds{R}}

\newcommand{\dist}{\stackrel{\mbox{\footnotesize d}}{=}}

 \newcommand{\zi}{z}

\newcommand{\xwrosA}{$\begin{array}{c}\vspace{-2ex}\\}
\newcommand{\xwrosB}{\\ \vspace{-2ex}\end{array}$}

\newcommand{\bbb}[1]{\mbox{\boldmath $ #1 $}}

 \title{ \Large\bf Self-Inverse
 and Exchangeable Random Variables\footnote{Work
 partially supported by the University of
 Athens Research Grant 70/4/5637.}
 \vspace{-.6em}}

 \author{\large
 Theophilos Cacoullos\footnote{
 e-mail:\ {\tt tkakoul@math.uoa.gr}}
 \ \ and \ \
 Nickos
 Papadatos\footnote{{\it Corresponding author.}
 e-mail:\ {\tt  npapadat@math.uoa.gr}
 \ \ \ url:\ {\tt http://users.uoa.gr/ $\sim$npapadat/}}}

 \date{\small
 Section of Statistics and O.R.,
 Department of Mathematics,
 University of Athens,
 \\ Panepistemiopolis, 157 84 Athens, Greece.
 }

\begin{document}

 \maketitle
 \vspace*{-2em}

 \thispagestyle{empty}

 \begin{abstract}
 \noindent
 A random variable $Z$ will be called self-inverse
 if it has the
 same distribution as its reciprocal $Z^{-1}$.
 It is shown that if $Z$ is defined as a ratio, $X/Y$,
 of two rv's $X$ and $Y$ (with $\Pr[X=0]=\Pr[Y=0]=0$),
 then $Z$ is self-inverse if and only if $X$ and $Y$ are (or can
 be chosen to be)
 exchangeable.
 In general, however, there may not exist iid $X$ and $Y$
 in the ratio representation
 of $Z$.
  \end{abstract}
 {\footnotesize {\it MSC}:  Primary 60E05.

 \vspace{-.7ex}
 variables;
 Exchangeable random variables;
 Representation of a self-inverse random variable as a ratio.}


 \section{Introduction}
 \vspace*{-1ex}
 \label{sec1}
 \setcounter{equation}{0}
 The definition of a self-inverse
 random variable (rv) is motivated by the observation that
 several known classical distributions are defined as the ratio
 of two independent and identically distributed
 (iid) rv's $X$ and $Y$, continuous as a rule, so that
 $\Pr[X=0]=\Pr[Y=0]=0$. Clearly, in this case $Z$ is self-inverse,
 that is,
 \be
 \label{eq.reciprocal}
 Z=\frac{X}{Y}\dist \frac{Y}{X}=Z^{-1},
 \ee
 where $X_1\dist X_2$ denotes that $X_1$ and $X_2$ have
 the same distribution.

 A classical example of a self-inverse
 rv $Z$ is the Cauchy with
 density
 \begin{eqnarray}
 \label{eq.Cauchy}
 f_Z(\zi)=\frac{1}{\pi}\ \frac{1}{1+\zi^2},  \ \ \ \zi\in\R,
 \end{eqnarray}
 since $Z$ is defined as the ratio of two iid
 $N(0,\sigma^2)$ rv's.
 The usual symmetry of $Z$,
 $Z\dist -Z$, is also obvious in (\ref{eq.Cauchy}).

 It may be added that not only such ratios of iid
 $N(0,\sigma^2)$ rv's have the Cauchy density;
 Laha (1958) showed that if $X$ and $Y$ are iid rv's
 with common density $f(x)=\sqrt{2}(1+x^4)^{-1}/\pi$,
 their  ratio also follows
 (\ref{eq.Cauchy}).
 In fact, interestingly enough, Jones (2008a) showed
 that
 the ratios
 $X/Y$ for all centered
 elliptically
 symmetrically distributed
 random vectors $(X,Y)$
 follow a general (relocated, $\mu\neq 0$,
 and rescaled, $\sigma\neq 1$) Cauchy,
 $C(\mu,\sigma)$.
 Such is the well-known case of a bivariate normal
 $(X,Y)$ with $X\sim N(0,\sigma^2)$ and
 $Y\sim N(0,\sigma^2)$; the ratio $Z=X/Y$ has
 the Cauchy density
 \begin{eqnarray}
 \label{eq.Cauchy2}
 f_Z(z)=\frac{1}{\pi}\
 \frac{\sqrt{1-\rho^2}}{1+z^2-2\rho z}
 =\frac{1}{\pi}\
 \frac{\sqrt{1-\rho^2}}{(1-\rho^2)+(z-\rho)^2},  \ \ \ z\in\R,
 \end{eqnarray}
 with $\mu=\rho$, the correlation
 coefficient, and scale parameter
 $\sigma=\sqrt{1-\rho^2}$.

 Arnold and Brockett (1992)
 showed that any random scale mixture of elliptically
 symmetric random vectors has a general
 Cauchy-type ratio (from any bivariate subvectors).
 Along the same lines, we add the very interesting article
 of Jones (1999),
 who used simple trigonometric formulas and
 polar coordinates to obtain Cauchy-distributed functions
 of spherically symmetrically distributed random vectors
 $(X,Y)$.


 In the present note we are not concerned with Cauchy-distributed
 ratios $X/Y$, known to be self-inverse, but with the question
 of when a random variable $Z$ has the same distribution as
 its reciprocal $Z^{-1}$, and of whether it is representable as a
 ratio $X/Y$. Seshadri (1965) considered the problem for a
 continuous rv $Z> 0$, and characterized the density
 $f_Z(z)$ of $Z$ in terms of the density $f_W(w)$ of $W=\log Z$:
 $f_W$ should be symmetric about the origin. This coincides
 with what Jones (2008b) refers to as ``log-symmetry'' about
 $\theta>0$:
 \[
 Z/\theta\dist \theta/Z;
 \]
 cf.\
 the so called ``$R$-symmetry'', introduced by
 Mudholkar and Wang (2007).
 Thus, our ``self-inverse'' symmetry for $Z>0$
 coincides with log-symmetry about $\theta=1$.
 Moreover, Seshadri (1965) showed that if $X$ and $Y$
 are iid, then $Z=X/Y$ is self-inverse; he also pointed out that
 the ratio decomposition of $Z$ into iid $X$ and $Y$ is not always
 possible. As already stated, we show below (Propositions \ref{prop1}
 and \ref{prop2})
 that the ratio representation of any self-inverse $Z$ is always
 possible in terms of two exchangeable rv's $X$ and $Y$. Also,
 two simple examples, showing that $X$ and $Y$ cannot always be chosen
 to be iid rv's, are given at the end of Section \ref{sec3}.

 \section{Examples of identically distributed rv's
 whose ratio is not self-inverse}
 \vspace*{-1ex}
 \label{sec2}
 Ratios $X/Y$ leading to (\ref{eq.Cauchy}) or the $F$-distributed
 $F_{n,n}\dist F_{n,n}^{-1}$, where $X$ and $Y$ are iid, are clearly
 self-inverse. In (\ref{eq.Cauchy2}), however, $X$ and $Y$ have the
 same distribution, but are not independent. One may, therefore,
 wrongly conclude that equidistribution of $X$ and $Y$ is a sufficient
 condition for the ratio to be self-inverse. This is not so, as
 shown by the following two examples, one discrete and one
 continuous. That $Z$ in (\ref{eq.Cauchy2}) is self-inverse is due
 to the fact that $X$ and $Y$, though not iid, are exchangeable.
 In Section \ref{sec3}, below,
 we show that
 the exchangeability of $X$ and $Y$
 is all we need to
 characterize
 a ratio $X/Y$
 as self-inverse.
 \smallskip


 \noindent
 {\small\bf (a) Discrete ($\bbb{X}$,$\bbb{Y}$).}
 The following table gives
 $f(x,y)=\Pr[X=x,Y=y]$:
 \begin{eqnarray}
 \label{eq.discrete}
 \mbox{
 \begin{tabular}{|cc|ccc|c|}
  \hline
  & $y$\vspace*{-.5em} & 1 & 2 & 3 & \\ 
  $x$ &  &  &  &  & $f_X(x)$\\ \hline
  1 &  & 2/36 & 9/36 &  1/36 & 1/3\\  
  2 &  & 1/36  &  2/36 &  9/36 & 1/3\\  
  3 &  & 9/36 & 1/36  &  2/36 &1/3 \\  \hline
    $f_Y(y)$ &  & 1/3 & 1/3 & 1/3 &1 \\
  \hline
 \end{tabular}
 }
 \end{eqnarray}
 Clearly, $X\dist Y$, with $X\sim U(\{1,2,3\})$,
 uniform on $\{1,2,3\}$. Yet, (\ref{eq.reciprocal})
 does not hold, since, for example,
 \[
 \Pr\bigg[\frac{X}{Y}=2\bigg]=\frac{1}{36}\neq
 \Pr\bigg[\frac{Y}{X}=2\bigg]=\frac{9}{36}.
 \]
 \smallskip

 \noindent
 {\small\bf (b) Continuous ($\bbb{X}$,$\bbb{Y}$)}.
 Let $U_1,U_2$ iid $U(0,1)$, i.e., uniform on $(0,1)$,
 and $I$ uniform on $\{0,1,2\}$, independent of $(U_1,U_2)$.
 Define $J=I+1$ if $I=0$ or $I=1$, and
 $J=0$ if $I=2$, so that $J\sim U(\{0,1,2\})$, that is,
 $J\dist I$. Observing that $(I,J)$ and
 $(U_1,U_2)$ are independent, and defining
 \[
 (X,Y)=(I+U_1,J+U_2),
 \]
 we have $X\overset{d}{=}Y\sim U(0,3)$, uniform on $(0,3)$.
 The joint density $f(x,y)$ of $X$
 and $Y$ is
 \begin{eqnarray}
 \label{eq.continuous}
  f(x,y)=\left\{\begin{array}{ll}
  1/3, & \text{if} \ \  x\in(0,1) \ \  \text{and} \ \  y\in(1,2), \\
  1/3, & \text{if} \ \  x\in(1,2) \ \  \text{and} \ \  y\in(2,3), \\
  1/3, & \text{if} \ \  x\in(2,3) \ \  \text{and} \ \  y\in(0,1), \\
  0,   &  \text{otherwise}.
 \end{array}\right.
 \end{eqnarray}
 Yet, the ratios $X/Y$ and $Y/X$ do not have the same
 distribution ((\ref{eq.reciprocal}) does not hold), since,
 e.g.,
 \[
 \Pr\bigg[\frac{X}{Y}\le1\bigg]=\frac{2}{3}, \ \
 \Pr\bigg[\frac{Y}{X}\le1\bigg]=\frac{1}{3}.
 \]
 In this example too, though $X\dist Y$, in fact $U(0,3)$,
 again (\ref{eq.reciprocal}) does not hold and $Z=X/Y$ is
 not self-inverse.
 \smallskip

 In both examples of (\ref{eq.discrete}) and
 (\ref{eq.continuous}), $X$ and $Y$ have the
 same distribution, but they are not exchangeable,
 and (\ref{eq.reciprocal}) fails. However, if $X$ and $Y$ are
 iid they are also exchangeable, since $F_X=F_Y$ and by
 independence,
 \begin{eqnarray}
 \label{eq.exch}
 F_{X,Y}(x,y)=F_X(x)F_Y(y)=F_Y(x)F_X(y)=F_{Y,X}(x,y).
 \label{eq8}
 \end{eqnarray}
 Such were the cases of (\ref{eq.Cauchy}) and $F_{n,n}$,
 and (\ref{eq.reciprocal}) holds;
 this also holds in (\ref{eq.Cauchy2})
 where $X$ and $Y$ are exchangeable,
 i.e., $F_{X,Y}=F_{Y,X}$.

 \section{Representation of a self-inverse random variable as a ratio}
 \vspace*{-1ex}
 \label{sec3}
 We have seen that if $X$ and $Y$ are not exchangeable,
 (\ref{eq.reciprocal}) may not hold, that is, the ratio
 $Z=X/Y$ may not be self-inverse. Here it will be shown
 that $Z$ is self-inverse if and only if it can be defined,
 or represented, as a ratio of two exchangeable rv's $X$
 and $Y$.

  First we show

 \begin{PROP}
 \label{prop1}
 Let $Z$ be defined as a ratio of
 two exchangeable rv's
 $X$ and $Y$, i.e.\
 \begin{eqnarray}
 \label{eq.exch2}
 Z=\frac{X}{Y},  \ \ \ \text{where} \ \
 (X,Y)\dist(Y,X) \ \ \mbox{and} \ \ \Pr[X=0]=\Pr[Y=0]=0.
 \end{eqnarray}
 Then $Z$ is self-inverse, that is,
 \begin{eqnarray}
 \label{eq.si}
 Z=\frac{X}{Y}\dist\frac{Y}{X}=Z^{-1}.
 \end{eqnarray}
 \end{PROP}
 \noindent
 \begin{Proof}
 In the continuous case where $(X,Y)$ has a density
 $f_{X,Y}(x,y)$ we may use the elementary formula
 for the density of $Z=X/Y$:
 \begin{eqnarray}
 \label{eq.density}
 f_Z(z)=\int_{0}^{\infty}y f_{X,Y}(yz,y)dy-\int_{-\infty}^0 yf_{X,Y}(yz,y)dy.
 \end{eqnarray}
 But $X$ and $Y$ are exchangeable,
 hence $f_{X,Y}=f_{Y,X}$,
 and (\ref{eq.density}) can be written as
 \begin{eqnarray}
 \label{eq.density2}
 f_Z(z)=\int_{0}^{\infty}y f_{Y,X}(yz,y)dy
 -\int_{-\infty}^0 yf_{Y,X}(yz,y)dy,
 \end{eqnarray}
 whose right hand side is the density of
 $Y/X=Z^{-1}$. Hence, $\mbox{(\ref{eq.exch2})}\Rightarrow\mbox{(\ref{eq.si})}$.

 In the general case, (\ref{eq.si}) is implied by the fact that if  $X$ and $Y$ are
 exchangeable, then,
 for any (Borel) function $g:\R^2\rightarrow \R$,
 we have
 \begin{eqnarray}
 \label{eq.g}
 g(X,Y)\dist g(Y,X).
 \end{eqnarray}
 Hence, taking $g(x,y)=x/y$ (with the convention
 $g(x,y)=0$ if $xy=0$), (\ref{eq.exch2}) implies (\ref{eq.si}).
 \medskip
 $\Box$
 \end{Proof}

 We are now going to show that, roughly speaking,
 (\ref{eq.si}) implies  (\ref{eq.exch2}), or more accurately:
 \begin{PROP}
 \label{prop2}
 If $Z\dist Z^{-1}$, there are exchangeable rv's
 $X$ and $Y$ (with $\Pr[X=0]=\Pr[Y=0]=0$)
 such that $Z$ can be written as
 \begin{eqnarray}
 \label{eq.exch3}
 Z\dist\frac{X}{Y}.
 \end{eqnarray}
 \end{PROP}
 \noindent
 \begin{Proof}
  Consider the pair
 \begin{eqnarray}
 \label{eq.constr}
 (X,Y)=(W Z^I,W Z^{1-I})=(W[(1-I)+IZ],W[I+(1-I)Z]),
 \end{eqnarray}
 where $I$ denotes the symmetric Bernoulli, with
 $\Pr[I=0]=\Pr[I=1]=\frac{1}{2}$, $W$ any rv with
 $\Pr[W=0]=0$, e.g., $W\equiv 1$ or $W\sim N(\mu,\sigma^2)$,
 and $Z,I,W$ are independent.
 It can be shown (cf.\ (\ref{eq.g}), (\ref{eq.exch}))
 that
 \[
 (X,Y)\dist (Y,X) \ \  \mbox{and, obviously,} \ \ \frac{X}{Y}=\frac{Z^I}{Z^{1-I}}.
 \]
 Hence, for any $z$ we have
 \begin{align*}
 \Pr\bigg[\frac{X}{Y}\leq z\bigg]
 &=\frac{1}{2}\Pr\bigg[\frac{Z^I}{Z^{1-I}}\leq z \bigg|
 I=1\bigg]
 +\frac{1}{2}\Pr\bigg[\frac{Z^I}{Z^{1-I}}\leq z\bigg|I=0\bigg] \\
 &=\frac{1}{2}\Pr[Z\leq z]
 +\frac{1}{2}\Pr\bigg[\frac{1}{Z}\leq z\bigg]=\Pr[Z\le\zi],
 \end{align*}
 since, by hypothesis, $Z\dist Z^{-1}$. Hence,
 \[
 Z\dist\frac{X}{Y}, \ \mbox{ with } (X,Y)\dist (Y,X)
 \mbox{ as defined by (\ref{eq.constr})}.
 \]
 \medskip
 $\Box$
 \end{Proof}

 Another question is whether there exist not simply
 exchangeable rv's $X$ and $Y$ as in (\ref{eq.exch3}),
 but iid $X,Y$ so that every self-inverse $Z$ can be
 written as in (\ref{eq.exch3}). The answer is negative,
 as
 shown by the following counterexample:

 Let $Z>0$ with $\log Z\sim U(-1,1)$ and suppose there are iid rv's
 $X,Y$ such that $Z$ can be written as in (\ref{eq.exch3}).
 Then it would follow that
 \begin{eqnarray}
 \label{eq.count2}
 \log Z=U\dist X_1-X_2 \ \ \text{with} \ \ X_1=\log|X|, \
 X_2=\log|Y|.
 \end{eqnarray}
 Moreover, since $X$ and  $Y$ are iid, the $X_1,X_2$
 will also be iid,
 in which case, if $\varphi$ is the characteristic
 function of  $X_1,X_2$, we have
 \begin{eqnarray}
 \label{eq.atopo}
 \varphi_{X_1-X_2}(t)=\varphi(t)\varphi(-t)
 =\varphi(t)\overline{\varphi(t)}=|\varphi(t)
 |^2\ge0,
 \end{eqnarray}
 whereas the characteristic function of  $U$
 is $\varphi_U(t)=(\sin t)/t$, taking both positive
 and negative values.

 An analogous (simpler) counterexample is the following:
  Let $Z>0$ with $\log Z=U$, $U$ the Bernoulli
 $\Pr[U=-1]=\Pr[U=1]=\frac{1}{2}$. Then, similarly as in
 (\ref{eq.count2}) and (\ref{eq.atopo}),
 \[
 \varphi_{X_1-X_2}(t)=|\varphi(t)|^2
 \ \ \mbox{whereas}\
 \varphi_U(t)=\cos t.
 \vspace{1em}
 \]

 \noindent
 {\small\bf Acknowledgement.} We thank the referee
 for suggesting Seshadri (1965) as an additional reference, and
 for comments which improved the presentation of this note.

\end{document}